\documentstyle[11pt,newpasp,twoside,epsf]{article}
\reversemarginpar

\def\vcirc{$V_{\it circ}$}
\def\sigc{$\sigma_c$}
\def\mbh{$M_\bullet$}
\def\mdm{$M_{\it DM}$}
\def\kms{$\rm km\;s^{-1}$}  

\begin{document}
\title{The $\sigma_c-V_{\it circ}$ correlation 
in high and low surface brightness galaxies}

\author{A. Pizzella, E. Dalla Bont\`a, E. M. Corsini, L. Coccato, and
F. Bertola} 
\affil{Dipartimento di Astronomia, Universit\`a di Padova,
Padova, Italy}

\begin{abstract}
We investigate the relation between the central velocity dispersion,
\sigc , and the circular velocity, \vcirc , in galaxies. 
In addition to previously obtained data, we consider an
observationally homogeneus sample of 52 high surface brightness and 11
low surface brightness spiral galaxies.
We performed a straight line regression analysis in a linear scale,
finding a good fit, also for low \sigc\ galaxies, always rejected
in the previous studies.
Low surface brightness galaxies seem to behave differently, showing
either higher values of \vcirc\ or lower values of \sigc\ with respect
to their high surface brightness counterparts.
\end{abstract}

\noindent
Studying the interplay between ionized-gas and stellar kinematics
allows to address different issues concerning the dynamical structure
of disk galaxies, and to constrain the processes leading to their
formation and evolution.
All these issues will greatly benefit from a survey devoted to the
comparative measurement of ionized-gas and stellar kinematics in disk
galaxies. This is particularly true for the relationships involving
the measurement of \sigc , such as the \mbh--\sigc\ relation
(Ferrarese \& Merrit 2000; Gebhardt et al. 2000) and the
\sigc--\vcirc\ relation (Ferrarese 2002; Baes et al. 2003).
To this aim we obtained the stellar and gaseous kinematics for 52
lenticular (Bertola et al. 1995) and spiral galaxies (Corsini et
al. 1999, 2003; Vega Beltr\'an et al. 2001; Pizzella et al. 2003)
of high surface brightness (HSB hereafter).
Recently, we considered a sample of 11 low surface brightness spirals
(LSB herafter). We obtained long-slit spectra along their major axes
at the Very Large Telescope with a resolution of about 45 km
s$^{-1}\times1''$. The ionized-gas kinematics was measured from the
H$\beta$ and [O~III]$\lambda5007$ emission lines. The stellar
kinematics was measured in the region of the Mg triplet at 5200 \AA\
with the Fourier Correlation Quotient method (Pizzella et al., these
proceedings).

For 42 sample galaxies we derived \sigc\ from the radial profile
of the stellar velocity dispersion and \vcirc\ from the flat portion
of the ionized-gas rotation curve (Fig. 1, left panel).  Finally, we
selected the 6 HSBs with data extending at radii larger than $R_{25}$
(and therefore ensuring a most reliable value of \vcirc ) and all the
11 LSBs. We built the final data set including the 24 HSBs studied by
Ferrarese (2002) and Baes et al. (2003) and 20 ellipticals studied by
Kronawitter et al. (2000).  We performed a straight line regression
analysis in a linear scale, finding a good fit, also for galaxies with
\sigc$<70$ \kms , always rejected in the previous studies (Fig. 1,
right panel).  Although our LSBs were not entirely representative of
the LSBs population, the comparison with the HSB sample seems to
indicate that LSBs follow a different relation, having a smaller
\sigc\ for a given \vcirc , or a higher \vcirc\ for a given \sigc .
The collapse of baryonic matter has been claimed to induce a further
concentration in the dark matter distribution, and a deepening of the
overall gravitational well in the central regions. If this is the
case, the finding that at a given \vcirc\ (which corresponds to a
given \mdm , Bullock et al. 2001) the central \sigc\ of LSBs is
smaller than in their HSB counterparts, would argue against the
relevance of baryon collapse in the radial density profile of dark
matter in LSBs.

\begin{figure}
\plottwo{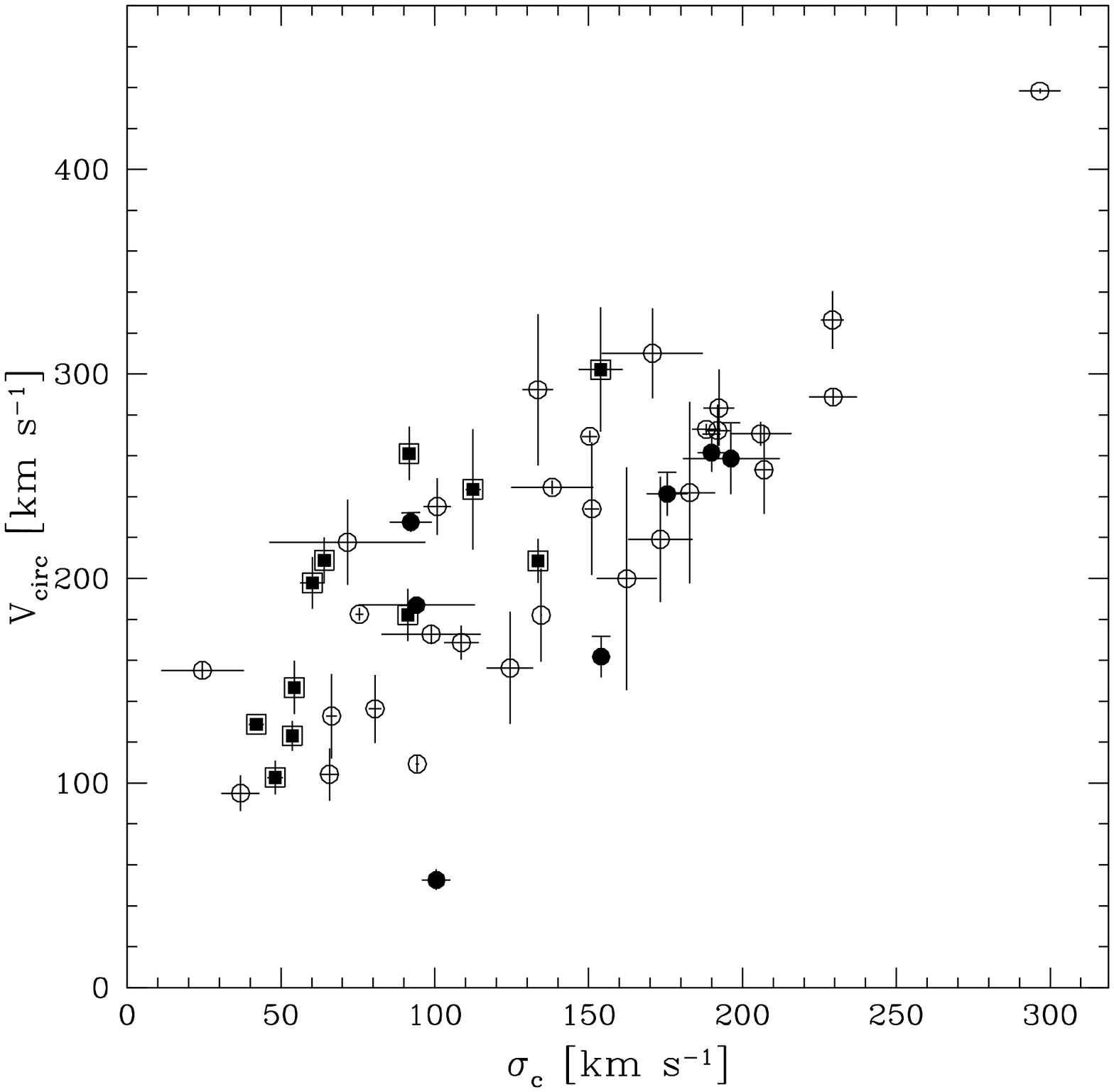}{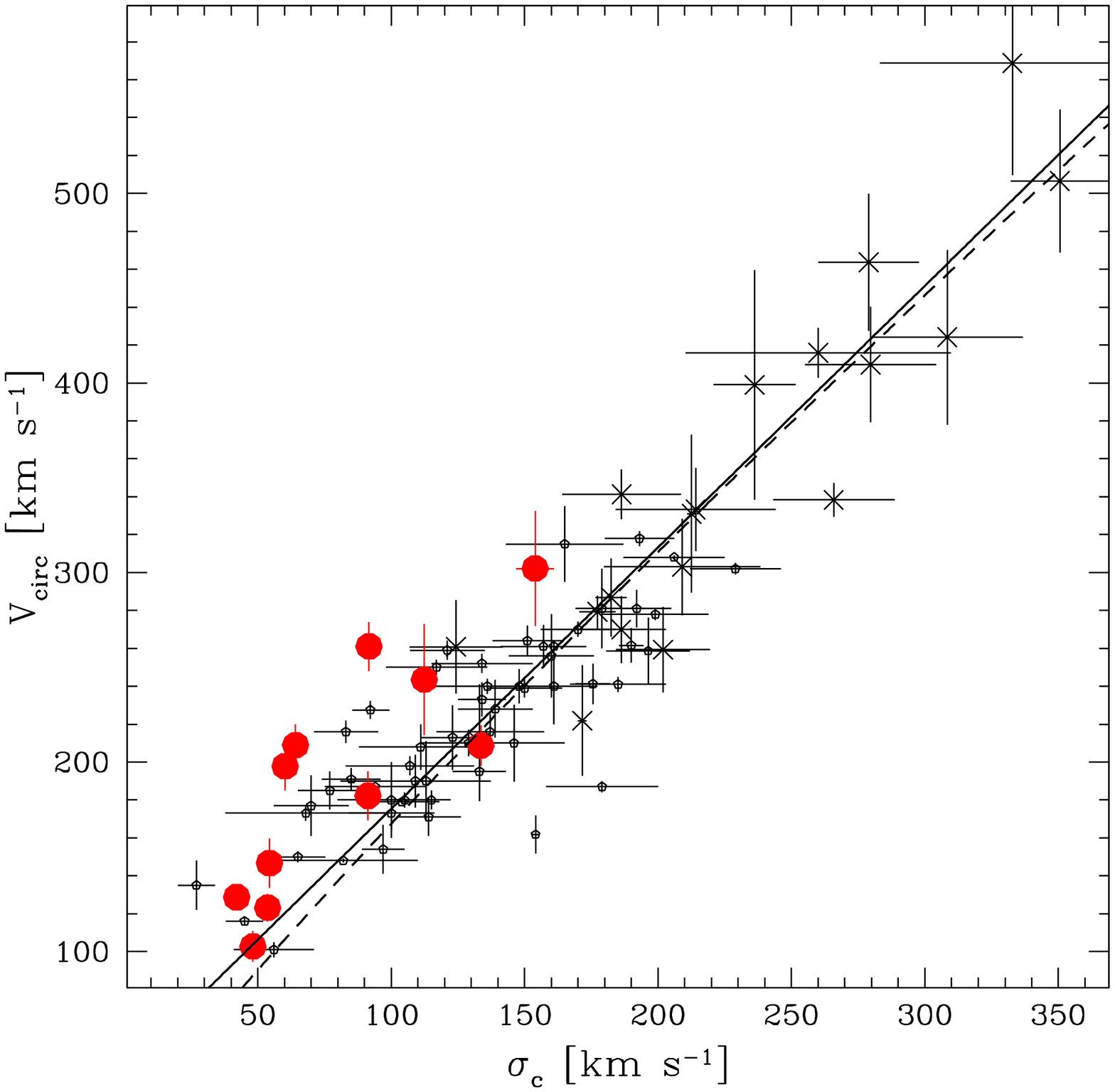}
\caption{{\it Left panel\/:} Values of \sigc\ and \vcirc\ for the 42
galaxies of our sample. {\it Open\/} and {\it filled circles\/}
correspond to HSBs for which $R(V_{\it circ})<R_{25}$ and $R(V_{\it
circ})>R_{25}$, respectively. LSB are marked with {\it squares\/}.
{\it Right panel\/:} \sigc--\vcirc\ relation for our LSBs ({\it
circles\/}), the HSBs with $R(V_{\it circ})>R_{25}$ taken from
Ferrarese (2002), Baes et al. (2003) and this paper ({\it dots\/}) and
the ellipticals of Kronawitter et al. (2000, {\it crosses\/}).  {\it
Continuous\/} and {\it dashed lines\/} correspond to our linear fit
and power-law fit by Ferrarese (2002), respectively.}
\end{figure}

\end{document}